\documentclass[prl,showpacs,twocolumn]{revtex4}

\begin{document}
\title{Ginsparg-Wilson Relation, 
           Topological Invariants \\
       and Finite Noncommutative Geometry }

\author{Hajime Aoki}
\affiliation{Department of Physics, Saga University, Saga 840-8502,
Japan; \\
e-mail: haoki@cc.saga-u.ac.jp
}

\author{Satoshi Iso and Keiichi Nagao}
\affiliation{
High Energy Accelerator Research Organization (KEK),
Tsukuba 305-0801, Japan; \\
e-mail: satoshi.iso@kek.jp, nagao@post.kek.jp
}

\date{\today}

\begin{abstract}
We show that the Ginsparg-Wilson (GW) relation can play an important role
to define chiral structures in {\it finite} noncommutative geometries.
Employing GW relation, we can prove the index theorem and construct
topological invariants even if the system has only finite degrees 
of freedom. As an example, we consider
a gauge theory on a fuzzy two-sphere and give an
explicit construction of a noncommutative analog of the
GW relation, chirality operator and the index theorem.
The topological invariant is shown to coincide with the 
1st Chern class in the commutative limit.
\end{abstract}


\maketitle

\setcounter{footnote}{0}
\paragraph{{\bf Introduction}} 
Quantization of space-time is one of
the unsolved issues  and lots of attempts have
been made. Superstring theory has succeeded in quantizing small
fluctuations of the metric,
but only little knowledge has been obtained for the space-time itself.
Discovery of D-branes has shown us a possibility that 
space-time coordinates can be noncommutative\cite{Witten}.
Based on this, various matrix models have been 
proposed as nonperturbative formulations of 
superstrings\cite{BFSS,IKKT,DVV}.
Furthermore, noncommutative geometries appear naturally 
in superstring theories with $B_{\mu \nu}$ 
background\cite{SeibergWitten} or in the matrix 
models\cite{CDS,AIIKKT}.
These studies have linked the string theory to the old
ideas that the concept of the space-time must be modified
at the Planck length\cite{Connes,tam}.
In order to understand the origin of our 
four dimensional space-time\cite{spacetime}, 
we need to analyze 
superstring numerically, and consider 
noncommutative geometries with 
{\it finite} degrees of freedom.

Another issue is the unification of space-time and the matter.
In particular, chiral structures of the fermions may 
be related to the quantization of space-time as we see in 
Penrose's twister theory or Connes' noncommutative geometry.
If we believe such connection, it is important to construct
chiral structures in systems with finite degrees of freedom.
This problem is also relevant to a construction of  four-dimensional
chiral fermions in matrix models.
Orbifolding\cite{AIS} or compactifications 
in matrix models with a nontrivial index 
can be candidates to define chiral structures in finite 
noncommutative geometries.
Instantons have been constructed
on noncommutative $R^4$ by using the shift 
operator\cite{instanton,harvey}, 
whose construction  essentially makes use of the infinite 
dimensionality of the Hilbert space.
In finite noncommutative theories, topologically nontrivial 
configurations have been constructed based on algebraic 
K-theory and projective modules\cite{Connes,balatop}. 
It seems difficult, however, to apply these ideas to finite 
square matrix models such as \cite{IKKT}.
\par
In this letter, we define a chirality and a Dirac
operator satisfying the Ginsparg-Wilson (GW) relation, which 
has been studied in the lattice gauge theory.
By using these operators, we can define  topological invariants
and prove an index theorem for general gauge field 
configurations, which are constructed from finite square 
matrices.
We first develop a general formalism and then give an explicit
example in the case of noncommutative (fuzzy) 2-sphere.

\paragraph{{\bf General Formalism}} 
In Connes' formulation of noncommutative geometry, 
a chirality operator and a Dirac operator which anti-commute 
are introduced.
In this letter, we propose to generalize 
this algebraic relation to the GW relation so that we can 
define chiral structures 
in a finite system representing a compact noncommutative manifold.
\par
GW relation has been studied to understand the chiral structure
in the lattice gauge theory with  a finite number of degrees of freedom.
It is expressed as 
\begin{equation}
D\Gamma_{n+1} + \Gamma_{n+1} D =a  D \Gamma_{n+1} D,
\label{GWintro}
\end{equation}
where $D$, $\Gamma_{n+1}$ and $a$ are  an $n$ dimensional 
lattice Dirac operator, a chirality operator and a lattice spacing
respectively.
This relation expresses the remnant chiral symmetry 
on the lattice, which was derived by  
the block-spin transformation
from a continuum theory\cite{GinspargWilson}. 
This algebraic relation makes it possible to define an extended chiral 
symmetry on the lattice\cite{Luscher,Nieder}. 
The index theorem  is also realized on the
lattice\cite{Hasenfratzindex,Luscher}.
Based on the idea of the domain wall fermion, a practical 
solution to the GW relation was obtained\cite{Neuberger}. 
Chiral gauge theory can be  
constructed only for an abelian case\cite{abeliangauge}. 
Higher dimensional generalization of the domain wall 
fermion was proposed toward constructing nonabelian 
chiral gauge theories\cite{KN}.
\par
Now we apply the algebraic structure of the GW relation to
finite noncommutative geometries.
In this context, the GW relation 
was first pointed out in refs.\cite{balaGW}
in connection to the fermion doubling problems.
The authors discussed the GW relation for free fermions 
on the fuzzy 2-sphere.
In this letter, by generalizing this result to 
fermions in general background gauge fields,
we can obtain topological invariants 
in finite noncommutative geometries.
GW relation is the most useful when it is applied to 
fermions coupled to general gauge field configurations.
We also note that on the noncommutative torus 
Dirac operator satisfying GW relation in a background 
gauge field is constructed 
in the paper~\cite{Nishi} to formulate chiral gauge theories.
\par
Our proposal here is that the GW relation can be 
substituted for the simple anticommutation relation, 
to define chiral structures for {\it general gauge fields} 
in general finite noncommutative geometries.
\par
In order to develop a general formalism, we 
consider a system that are composed
of $N \times N$ matrices. $1/N$ plays a role of the lattice
spacing $a$ in the lattice gauge theory.
We assume an existence of a chirality
operator  $\gamma$  and a Dirac operator in the commutative limit,
where $N$ goes to infinity (or $a$ goes to zero). 
We then  introduce two chirality operators $\Gamma$
and $\hat{\Gamma}$ and a Dirac operator $D$ in our finite system.
They are operators acting on $N \times N$ matrices. 
The Dirac operator $D$ generally depends on background gauge 
field configurations. Both chirality operators 
are required to become the chirality operator $\gamma$
in the commutative limit and satisfy the relations:
\begin{equation}
\Gamma^2=\hat{\Gamma}^2=1, \ \ \Gamma^{\dagger}=\Gamma, \ 
\hat{\Gamma}^{\dagger} = \hat{\Gamma}.
\label{gamma}
\end{equation}
$\Gamma$ is assumed to be 
independent of the gauge field configurations.
If the system has chiral anomaly in the commutative limit,
we cannot expect that the Dirac operator $D$
anti-commutes with the chirality operator $\Gamma$,
since, if so, it will contradict with the result in the commutative limit.
Hence, in the finite noncommutative geometry, a simple 
algebraic structure is destroyed and we cannot prove an index 
theorem only with the simple chirality operator $\Gamma$.
\par
The other chirality operator $\hat{\Gamma}$ can be constructed in terms of
a hermitian operator $H$ as 
\begin{equation}
\hat{\Gamma} \equiv \frac{H}{\sqrt{H^2}}, \ H^{\dagger}=H.
\label{hatgamma}
\end{equation}
$\hat\Gamma$ depends on the gauge field configuration 
through $H$.
\par
If we take an appropriate choice of $H$, we can define a new
Dirac operator (GW Dirac operator) $D_{GW}$ by
\begin{equation}
1- \Gamma \hat{\Gamma} = f(a,\Gamma) D_{GW}.
\label{GWdef}
\end{equation}
The prefactor $f$ is assumed to be a function of 
the small parameter $a$ and the chirality operator $\Gamma$. 
These $f$ and $H$ are determined so that the Dirac operator 
$D_{GW}$ becomes the commutative Dirac 
operator in the limit $a \rightarrow 0$.
From (\ref{gamma}) and (\ref{GWdef}), we have the relation
\begin{equation}
\Gamma D_{GW}+D_{GW} \hat{\Gamma}=0,
\label{GWnew}
\end{equation}
and then we can prove the following index theorem:
\begin{equation}
2 \ {\text {index}}D_{GW} \equiv 2(n_+ - n_-) = 
{\cal T}r (\Gamma + \hat{\Gamma}),
\label{indexth}
\end{equation}
where ${\cal T}r$ is a trace of operators acting on matrices, 
and $n_{\pm}$ are numbers of zero eigenstates of $D_{GW}$
with a positive (or negative) chirality (for either $\Gamma$
or $\hat{\Gamma}$).
\par
It is easy to prove that this index is invariant
under a small deformation of any parameter or any
configuration (such as gauge field) in the operator $H$.
A necessary check is that this index can take a nontrivial
value depending on the background configuration.
For some choices of $H$, this index may become a constant
and trivial. For other choices, it gives a nontrivial 
index as we will see in an example on fuzzy 2-sphere.
This is surprising since it seems impossible to define such a
nontrivial index out of finite matrices. The resolution lies in the
definition of the operator $\hat{\Gamma}$ in  (\ref{hatgamma}). 
It becomes singular when the operator $H$ has a zero-mode.
It is a sign function of the operator $H$,
and when an eigenvalue of $H$ crosses zero the index changes
by two. Therefore, if there is no symmetry among eigenvalues, 
the index can take nontrivial values.
The configuration space of gauge fields are divided into
islands, in each of which the index takes a  different value.
In the lattice gauge theory, we can exclude a region where
$H$ has zero eigenvalues by imposing an admissibility condition
for the gauge field\cite{locality} . 
We expect a similar condition for our case.
It is also interesting to investigate these conditions 
and the topological properties for the GW Dirac operator 
on noncommutative torus constructed in the paper~\cite{Nishi} 
since it has closer relationship to the ordinary lattice 
gauge theory. 
\par
Another necessary check is to reproduce the well-known index 
theorem in the commutative limit. Namely, the topological 
invariant considered here should give an instanton number 
for nontrivial gauge field configurations.
In the case of fuzzy 2-sphere, we will show that the index 
reproduces the monopole number.
\par
Regarding this commutative limit, it is amusing to point out 
that we will be able to arrive at different commutative limits 
with different types of topological invariants
though we start from the same size of matrices.
Namely, topological invariants in different space-time dimensions 
can be obtained from the same matrices.
Different classical interpretations come from different choices 
of chirality
operators and  Dirac operators, which 
 lead to different embeddings of 
the commutative configurations in the matrices.
In refs.\cite{Kiskis,KikukawaSuzuki} they
gave simple examples of the embeddings of classical configurations
in matrices and obtained classical indices on the lattice.
Since an important property of the noncommutative theories
is that it contains much more degrees of freedom than ordinary
field theories, 
it is interesting to investigate if we can classify the universality classes
of more general embeddings in the matrices.
We want to discuss the problem in the future\cite{AINfuture}. 
\paragraph{{\bf Chiral Transformation}}
We can construct a chiral invariant model 
by using the Dirac operator $D_{GW}$ as
\begin{equation}
 S= {\text tr}(\bar\Psi D_{GW} \Psi).
\end{equation}
Each component of $\Psi$ is an $N \times N$ matrix
and ${\text tr}$ means a trace over matrices.
This action is invariant under the global chiral 
transformation:
$\delta\Psi=i \hat{\Gamma}\Psi,\delta\bar\Psi=i \bar\Psi \Gamma.$
To make it local, we need to specify the 
ordering of the chiral transformation parameter $\lambda$
and the fermion field.
The fermion transforms as
 the fundamental representation under gauge transformations:
\begin{equation}
\Psi \rightarrow U \Psi, \ \bar\Psi \rightarrow \bar\Psi U^{\dagger}.
\label{gaugetr}
\end{equation}
$D_{GW}$ must be constructed to maintain the gauge invariance.
Namely we require the transformation, 
$D_{GW} \Psi \rightarrow U D_{GW} \Psi$ 
under gauge transformations.
Then, if we put $\lambda$ in the left of $\Psi$, 
$\lambda$ should transform covariantly as
$\lambda \rightarrow U \lambda U^{\dagger}$ and 
so does the chiral current.
On the contrary, if we put $\lambda$ in the right,
$\lambda$ and the associated chiral current is invariant
under gauge transformations.
This ambiguity is specific to the noncommutative field theories 
and makes the analysis of the Ward-Takahashi(WT)
identity complicated.
This issue was discussed in refs.\cite{anomaly,AIN}.
In this letter, we don't go further into this problem and
restrict our discussion to the covariant case.
\par
In obtaining the WT identity for the covariant
chiral current, the integral measure of the fermion fields
is not invariant under the local chiral transformations,
\begin{equation}
 \delta \Psi =i \lambda \hat{\Gamma} \Psi, \ \
 \delta \bar{\Psi} = i \bar{\Psi}\lambda \Gamma,
\end{equation}
and produces a nontrivial Jacobian.
This term gives an integral of the 
 topological charge density (an anomaly term)
with the local weight $\lambda$
in the WT identity:
\begin{equation}
q(\lambda)=\frac{1}{2}{\cal T}r(\lambda^L \hat{\Gamma} +\lambda^R \Gamma')
= \frac{1}{2}{\cal T}r(\lambda^L \hat{\Gamma} +\lambda^L \Gamma).
\label{topchargedensity}
\end{equation}
Here the superscript $L$ ($R$) in $\lambda$ means that this
operator acts from the left (right) on matrices.
$\Gamma'$ is obtained from $\Gamma$ 
by exchanging $L$ and $R$ superscript.
For a global chiral transformation, we set $\lambda=1$ and
$q(\lambda)$ becomes the index defined in (\ref{indexth}).
\paragraph{ {\bf Examples on Fuzzy 2-Sphere}}
We now consider a simple example on fuzzy 2-sphere.
Noncommutative coordinates of the 
fuzzy 2-sphere is given by
$ x_i =\alpha L_i$,
where $L_i$'s are  $2L+1$-dimensional irreducible 
representation matrices of $SU(2)$ algebra.
The radius of the sphere is given by $\rho=\alpha \sqrt{L(L+1)}$.
Wave functions on fuzzy 2-sphere can be expanded in terms of noncommutative
analogs of the spherical harmonics $\hat{Y}_{lm}$.
They are traceless symmetric products of
the noncommutative coordinates.
There is an upper bound for the angular momentum
$l$ for $\hat{Y}_{l,m}$; $l \le 2L$.
Any hermitian matrix $M$ can be expanded in terms of 
these spherical harmonics $\hat{Y}_{l,m}$.
\par
Killing vectors on the fuzzy 2-sphere are expressed 
by taking a commutator with the $SU(2)$ generator; 
$ {\cal L}_i M= [L_i, M] =(L_i^L-L_i^R)M. $
An integral over 2-sphere is replaced by taking a trace over matrices
$ \frac{1}{2L+1} {\text tr} \leftrightarrow \int 
\frac{d \Omega}{4 \pi}. $
More detailed correspondence are found in refs.\cite{IKTW,AIN}.
\par
Now we introduce a Dirac operator $D$ and a chirality operator
$\Gamma^R$ as follows: 
\begin{eqnarray}
D&=&\sigma_i({\cal L}_i+\rho a_i^L)+1,  \label{Diracaction} \\
\Gamma^R &=& \frac{1}{2L+1}(2\sigma_i L_i^R -1).
\label{GammaR}
\end{eqnarray}
The free part of $D$ contains no fermion 
doublers\cite{balagovi}.
This $\Gamma^R$ satisfies the conditions (\ref{gamma}).
In the commutative limit, it
 reduces to an ordinary chiral operator on the sphere
$\gamma=\sigma_i x_i/ \rho$.
An anticommutator with the Dirac operator (\ref{Diracaction}) 
becomes
\begin{equation}
\{\Gamma^R,D \}= \frac{1}{L+1/2} \left(
2({\cal L}_i+\rho a_i^L)L_i^R - D
\right). \label{gammaDac}
\end{equation}
Here ${\cal L}_i L_i^R$ and $D/(L+1/2)$ vanish in the 
commutative 
limit since they are of order $1/L$.
Hence the anti-commutator becomes proportional to the 
scalar field:
\begin{equation}
\{\Gamma^R,D \} \rightarrow 2  a_i x_i = 2 \rho \phi.
\end{equation}
Note that since we embed the 2-sphere in three dimensional
flat space, the normal component of the gauge field $a_i$
is interpreted as a scalar field on 2-sphere.
\par
We then introduce
the other chirality  operator $\hat{\Gamma}$ as in (\ref{hatgamma}).
We construct $H$ out of $\Gamma_R$ and $D$
because these operators commute with the gauge transformations
(\ref{gaugetr}). This condition is required from the 
compatibility condition for chiral transformations generated by
$\hat{\Gamma}$ and gauge transformations. 
If we restrict 
$H$ to contain $D$ less than 2, it becomes 
\begin{equation}
H= \Gamma^R + c_1 D+ ic_2 [D, \Gamma^R]
+ c_3 \left\{\Gamma^R,D \right\} +c_4 \Gamma^R D  \Gamma^R +c_5. 
\end{equation}
The coefficients $c_i$ are real and of order $1/L$ 
so that $\hat\Gamma$ becomes $\gamma$ in the commutative limit.
The next requirement is that, if we define 
$f$ in (\ref{GWdef}) appropriately,
$D_{GW}$ must become an ordinary Dirac operator
in the commutative limit.
Indeed, if we choose $f=(c_4-c_1)\Gamma^R +2 i c_2$, 
the GW Dirac operator becomes
$D_{GW}=(D-\{\Gamma^R,D \}\Gamma^R/2)+{\cal O}(1/L)$.
It is nothing but the Dirac operator whose
coupling to the scalar is projected out.
The Dirac operator $D$ is coupled to three 
dimensional gauge fields $a_i$ and on the sphere
they are docomposed into two dimensional gauge field
and a scalar field.
Since the  GW Dirac operator satisfies the
GW relation (\ref{GWnew}) and anticommutes
with the chirality operator in the commutative
limit, it is natural that the coupling to
the scalar field is projected out from $D_{GW}$.
\par
In the following, for calculational simplicity, 
we take $H$ as,
\begin{equation}
H=\Gamma^R + a D = \Gamma^L +a \rho \sigma_i a_i^L,
\end{equation}
where $a=1/(L+1/2)$ and we defined $\Gamma^L$ 
as $\Gamma^L=(2\sigma_i L_i^L +1)/(2L+1)$.
This operator also satisfies $(\Gamma^L)^2=1$ and 
$(\Gamma^L)^{\dagger}=\Gamma^L$.
Since both of $\Gamma^L$ and $a_i^L$ are operators acting 
on matrices from the
left, this $H$ also acts only from the left.
Then the topological charge density (\ref{topchargedensity})
becomes
\begin{equation}
q(\lambda)=\frac{1}{2}{\text tr}(1)
{\text Tr} (\lambda \hat{\Gamma})
+ \frac{1}{2}{\text tr}(\lambda) {\text Tr}(\Gamma^{R}).
\end{equation}
${\text Tr}$ is a trace over matrices and spinors,
and $\Gamma^R$ and $\hat{\Gamma}$
are considered here as mere matrices instead of operators 
acting on matrices from the right and from the left respectively.
If there is no background gauge field, $\hat{\Gamma}=\Gamma^L$
and the topological charge density vanishes.
If we assume that the gauge field configuration is weak, 
we can expand the chirality operator $\hat{\Gamma}$ 
with respect to the gauge field $a_i$.
Up to the first order in $a_i$, after taking a trace over $\sigma$ matrices,
we obtain
\begin{equation}
q(\lambda)=-\frac{a^3 \rho (2L+1)}{2\alpha}  {\text tr} \lambda \left( 
i \epsilon_{ijk}[L_i, a_j]x_k +2\rho\phi -\frac{\alpha}{2}[L_i,a_i]
\right).
\end{equation}
In the commutative limit, this becomes
\begin{equation}
q(\lambda) = 2\rho \int  \frac{d \Omega}{4 \pi} 
\lambda \epsilon_{ijk} x_i \partial_j a'_k
\label{chern}
\end{equation}
where $a'_i$ is a tangential component of $a_i$;
$a'_i = {\epsilon_{ijk}x_j a_k / \rho }$. 
This topological charge density is nothing but the monopole
charge density on the 2-sphere.
For $\lambda=1$, this index is quantized to be integers.
If we embed classical gauge field configurations with
nontrivial monopole charge, our index gives a nontrivial value.
\paragraph{{\bf Discussions}}
In this letter we have proposed to use Ginsparg-Wilson 
relation to formulate the chiral and topological structures
in the finite noncommutative geometry, that is, in 
matrix models. 
As an example, we constructed a GW Dirac operator and
a GW chirality operator for fermions in background gauge field
configurations on fuzzy 2-shpere.
We then obtained a topological invariant for the gauge field
configurations. This invariant is shown to become the first Chern number
in the commutative limit if we assume that the gauge field is weak enough.
\par
As we will report soon in a separate paper, 
the invariant can take a nonzero integer
and hence this invariant is nontrivial.
This looks puzzling at first sight because the gauge field $a_i$
is defined globally in the matrix models even though the configuration
can have a nontrivial topology on two-sphere.
This problem will be solved if we consider patches on the
sphere and note that the gauge field configuration can be taken small 
only in a single patch. Hence our assumption in obtaining
(\ref{chern}) is only valid locally.
It will be interesting if we can construct a concept of
patches in noncommutative geometries.
\par
If we take a different $H$, we expect that the topological
charge has the same commutative limit (i.e. Chern number)
but there will be a slight difference 
before taking the commutative limit.
This problem is related to the issue of the universality
in the lattice gauge theories. It is an important problem
to clarify the 'universality' in the noncommutative gauge theories.
\par
There are  several other problems listed below.
\par
First the fermion takes a matrix form
and  it contains $N^2$ degrees of freedom.
Since the number of lattice points is
considered to be of order $N$, the fermion contains
much larger degrees of freedom than the ordinary field theories.
This is also true for the gauge field configuration.
These larger degrees of freedom have been interpreted 
as an indicator that the noncommutative field theory is 
related to string theory\cite{NCYM}. Furthermore, it has been
expected that space-time is also dynamical in the
noncommutative Yang-Mills theories.
In the commutative limit, 
we restrict wave functions to those with momentum smaller than
the noncommutativity scale and then the degrees of freedom
become of order $N$. 
Classical interpretations will be available only for this 
limiting cases.
Different embeddings of classical configurations
lead to different classical theories.
From this argument, our index is expected to classify 
even the global topology of space-time.
For this purpose,
it is important to investigate more examples on higher
dimensional noncommutative geometries.
\par
Another problem is the relation to  other arguments
on the topological properties of the noncommutative field 
theories\cite{Connes,harvey,balatop}. 
Topologically nontrivial configurations are constructed 
by projection operators. 
It will be interesting to investigate relations to them.
\par
Finally we want to comment on the admissibility condition
for the gauge field configurations.
The GW Dirac operator, or the chirality operator $\hat{\Gamma}$
become ill-defined when the operator $H^2$ has a zero eigenvalue.
In the lattice gauge theory, we can exclude such configurations
by imposing a condition that the field strength on each plaquette is 
smaller than a critical value\cite{locality}. 
This condition is called
an admissibility condition and the gauge field configurations
satisfying it are classified for an abelian gauge 
field\cite{abelian}. 
In our case, we can also expect a similar condition.
It will be reported in near future.
\par
We would like to thank T. Onogi, M. Peskin and H. Suzuki 
for discussions, and A. P. Balachandran for informing us 
of some references on topological properties based on the 
algebraic K-theory.

\end{document}